\documentclass[conference]{IEEEtran}
\usepackage{cite}
\ifCLASSINFOpdf
  \usepackage[pdftex]{graphicx}
\else
\fi
\usepackage{multirow}

\usepackage{amsmath}
\usepackage{algorithmic}
\usepackage{pbox}
\usepackage{array}
  \usepackage[caption=false,font=footnotesize]{subfig}
\begin{document}
%\title{Bare Demo of IEEEtran.cls\\ for IEEE Conferences}
\title{KidsTube: Detection, Characterization and Analysis of Child Unsafe Content \& Promoters on YouTube}
%\author{\IEEEauthorblockN{Michael Shell}
%\IEEEauthorblockA{School of Electrical and\\Computer Engineering\\
%Georgia Institute of Technology\\
%Atlanta, Georgia 30332--0250\\
%Email: http://www.michaelshell.org/contact.html}
%\and
%\IEEEauthorblockN{Homer Simpson}
%\IEEEauthorblockA{Twentieth Century Fox\\
%\and
%\IEEEauthorblockN{James Kirk\\ and Montgomery Scott}
%\IEEEauthorblockA{Starfleet Academy\\
%San Francisco, California 96678--2391\\


\author{\IEEEauthorblockN{Rishabh Kaushal\IEEEauthorrefmark{1},
Srishty Saha\IEEEauthorrefmark{1},
Payal Bajaj\IEEEauthorrefmark{2} and
Ponnurangam Kumaraguru\IEEEauthorrefmark{1}}\and\and
\IEEEauthorblockA{\hspace{2cm}\IEEEauthorrefmark{1} Cybersecurity Education and Research Centre (CERC) \\\hspace{2cm} Indraprastha Institute of Information Technology, Delhi. 
\\\hspace{2cm} Email: \{rishabhk, srishty12107, pk\}@iiitd.ac.in}\and
\IEEEauthorblockA{\IEEEauthorrefmark{2} Adobe Research \\
Bangalore, India. \\ Email: pabajaj@adobe.com}}


\maketitle

\begin{abstract}
\textit{YouTube} draws large number of users who contribute actively by uploading videos or commenting on existing videos. However, being a crowd sourced and large content pushed onto it, there is limited control over the content. This makes malicious users push content (videos and comments) which is inappropriate (unsafe), particularly when such content is placed \textit{around} cartoon videos which are typically watched by kids. In this paper, we focus on presence of unsafe content for children and users who promote it. For detection of child unsafe content and its promoters, we perform two approaches, one based on supervised classification which uses an extensive set of video-level, user-level and comment-level features and another based Convolutional Neural Network using video frames. Detection accuracy of 85.7\% is achieved which can be leveraged to build a system to provide a safe YouTube experience for kids. Through detailed characterization studies, we are able to successfully conclude that unsafe content promoters are less popular and engage less as compared with other users. Finally, using a network of unsafe content promoters and other users based on their engagements (likes, subscription and playlist addition) and other factors, we find that unsafe content is present very close to safe content and unsafe content promoters form very close knit communities with other users, thereby further increasing the likelihood of a child getting getting exposed to unsafe content. 
\end{abstract}

\IEEEpeerreviewmaketitle

\section{Introduction}
\begin{figure*}[!t]
\centering
\subfloat[]{\includegraphics[width=2.0in]{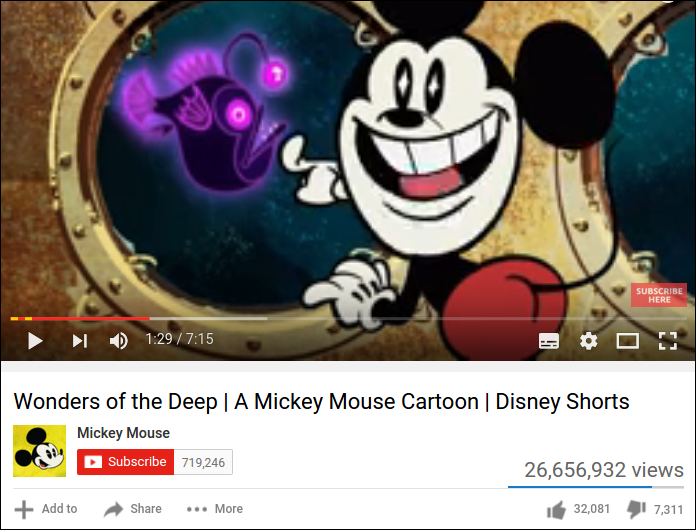}%
\label{fig:decent_cartoon}}
\hfil
\subfloat[]{\includegraphics[width=2.0in]{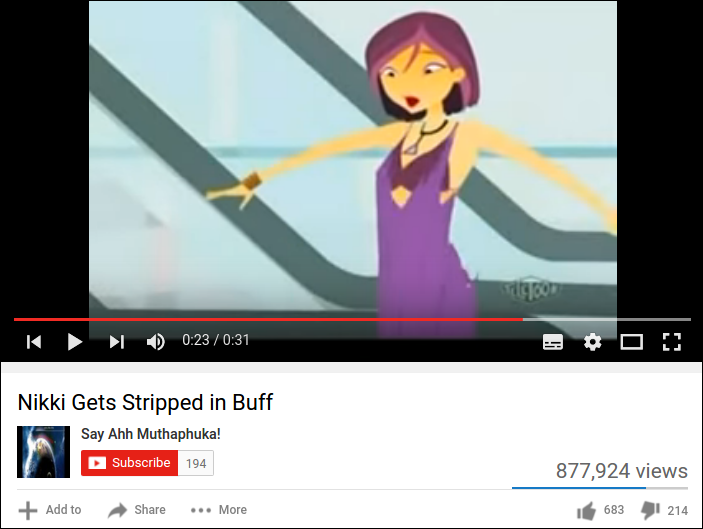}%
\label{fig:indecent_cartoon}}
\hfil
\subfloat[]{\includegraphics[width=2.0in]{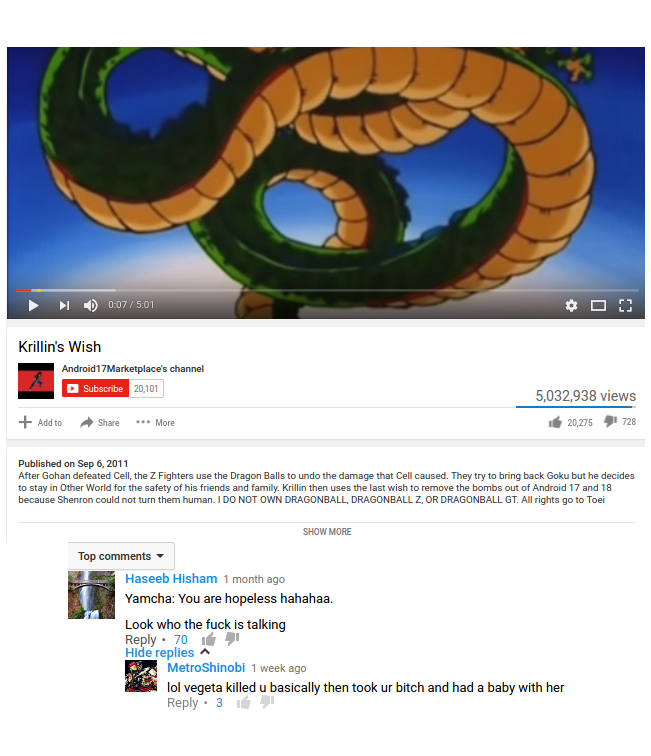}%
\label{fig:bad_comment}}
\caption{(a) A famous cartoon character `Mickey Mouse', example of cartoon video on YouTube which is safe for children, it is widely watched as indicated through the large view count of 26,656,932. (b) Example of cartoon video which depicts visual nudity and is unsafe for children, such videos can be found easily on YouTube. (c) Example of unsafe comments on otherwise safe video.}
\end{figure*}

YouTube is the most popular video sharing platform on the web. It enjoys a global $2^{nd}$ rank among websites viewed in the world as of today.\footnote{http://www.alexa.com/siteinfo/youtube.com} According to the statistics released by YouTube \cite{youtube-stats}, the user base of YouTube is over a billion users and 400 hrs of viewable content is getting uploaded in 60 seconds \cite{yt-stats}. YouTube provides ease of publishing and is highly leveraged by content producers to host their video content on YouTube. Consequently, a wide variety of content is posted on YouTube and watched by users across all age groups\cite{yt-demographics}. Our work focusses on cartoon videos which are typically watched by children. Various cartoon production houses have uploaded their trademark cartoon series on YouTube for wider publicity and brand building. Figure \ref{fig:decent_cartoon} depicts one such cartoon video which clearly indicate that such cartoon videos receive large view counts and thus are widely watched.  
%\begin{figure}[h]
%\centering
%\includegraphics[width=0.35\textwidth]{decent_cartoon1.png}
%\caption{A famous cartoon character `Mickey Mouse', an example of cartoon video on YouTube which is appropriate for children and is widely watched as indicated through the large view count of 26,656,932.}
%\label{fig:decent_cartoon}
%\end{figure}
This large audience, typically in the lower age group, has very less experience of nuances and subtleties that Internet has to offer. Being a crowd sourced platform, YouTube, is no exception to users with malicious intent who are often found indulging in spamming and promotional activities. In the context of these children, the most disruptive behaviour by malicious users is to expose them to inappropriate content, Figure \ref{fig:indecent_cartoon} is one such example. 
%\begin{figure}[h]
%\centering
%\includegraphics[width=0.35\textwidth]{indecent_cartoon1.png}
%\caption{An example of cartoon video which depicts visual nudity and is inappropriate for children, such videos can be found easily on YouTube.}
%\label{fig:indecent_cartoon}
%\end{figure}
Unsafe content on YouTube is contributed either in form of indecent videos published by \textit{unsafe uploaders} or in form of abusive/inappropriate words in comments post by \textit{unsafe commenters}. We refer such uploaders and commenters in our paper as \textit{child unsafe content promoters}. An otherwise safe video page can be rendered as inappropriate due to presence of abusive comments, as shown in Figure \ref{fig:bad_comment}. For the purpose of our work, indecent videos are those which either have graphic nudity or abusive/inappropriate dialogues. Table \ref{table:user_types} depicts all possible scenarios of publishing of unsafe content. 
\begin{table}[h!]
\centering
\caption{All possible combinations of User Behaviour, here `1' represents an Unsafe behaviour while `0' represents a Safe behaviour.}
\begin{tabular}{ |c|c|c|}\hline
\textbf{Video} & \textbf{Comment} & \textbf{User Behaviour}\\ \hline 
0 & 0 & Safe Uploads \& Safe Comments\\
0 & 1 & Safe Uploads \& Unsafe Comments\\
1 & 0 & Unsafe Uploads \& Safe Comments\\
1 & 1 & Unsafe Uploads \& Unsafe Comments\\
 \hline
\end{tabular}
\label{table:user_types}
\end{table}

\textbf{Motivation}: There have been a number of articles indicating the issue of unsafe content on YouTube, one of them claiming that kids are three clicks away from adult content on YouTube \cite{three-clicks} and another suggesting that children are at a higher risk of accessing adult content on YouTube \cite{children-high-risk}. YouTube tried to address these concerns through YouTube Kids App \cite{youtube-kids}. However, it was based on user feedback (viewer flagging unsafe content), which could be manipulated. This solution could not provide expected results and consumer groups reported to FTC that it still has inappropriate content \cite{app-inappropriate} which YouTube later assured would be addressed in future updates \cite{app-complaint}. 

\textbf{Key Contributions}: We summarize our contributions as three fold. \textit{First}, we provide two approaches to detect unsafe content promoters (Section IV). One uses machine learning based supervised classification model based on an extensive feature set at the level of video, user and comments. Second builds a Convolutional Neutral Network based deep learning model using features from the video frames. \textit{Second}, we perform a detailed comparative characterization study (Section V) of unsafe content promoter behaviour and engagements received by them with other users spreading safe content and find distinguishing patterns. \textit{Third}, using network analysis (Section VI), we show that transition of a YouTube user (typically a child in our case) watching a decent video to an indecent video is highly likely. Further, we model unsafe content promoters and other users as a graph built using their behaviours of likes, subscription and playlist. We then apply community detection algorithms to unravel closely knit communities of unsafe content promoters with normal users.
\section{Related Work}
We discuss some key indicative research problems and challenges on YouTube that have been investigated related to the problem under study.
\subsection{Detection of Inappropriate Content on YouTube}
Detection of inappropriate video content (pornographic) by combining image features with motion information was studied by Jansohn et. al. \cite{porn-detect} in which motion extracted in the form of MPEG-4 motion vectors was used. Ochoa et. al. \cite{adult-detect} used spatial and temporal cinematographic features from video to build machine learning classifiers to perform binary classification of videos into adult and non-offensive. In our work too we followed the path of using video frame and image processing based features in one of our approaches based on Convolutional Neural Network.
\subsection{Detection of malicious user behaviour}
Being a crowd sourced platform, users often are found to exhibit malicious behaviour on YouTube. One work to detect spammers and content promoters was done by Benevenuto et. al. \cite{fabricio} in which authors manually annotated a set of YouTube users as spammers, promoters and legitimate users. This was followed by characterization and use of classification algorithms for categorizing a user into these categories. Our work draws methodological insights from their work. However, our work is focussed on the issue of unsafe content promotion in respect to cartoon videos in contrast to their work which looks at spammers and promoters in general. Another closely related problem of identification of fraudulent promotion of videos was  addressed by Bulakh et. al. \cite{fraud-video} in which measurement studies were performed to distinguish fraudulent promoters from legitimate users which was followed by development of supervised machine learning models to differentiate legitimate users from those indulging in fraudulent promotion. Spam campaigns in YouTube is another interesting problem studied by Callaghan et. al. \cite{spam-campaign} who observed broadly two spam strategies. One strategy worked by small group of users commenting on large set of videos and another worked by large group of users commenting on small set of videos. Sureka et. al. \cite{sureka} investigated spamming activity in the comments posted on YouTube videos and presented an empirical analysis of comments obtained from YouTube and finally proposed a rule based classification method to detect spammers in comment activities. Our work also looks at comments but we do it from the point of view of detection of inappropriate words in it and study the relationship between commenters and uploaders of videos. Works of Agarwal et. al. \cite{crawler} and Sureka et. al. \cite{hidden} address the problem of mining and extracting hate and extremism videos on YouTube, but they were studied to understand the retrieval approaches being followed to obtain the desired set of videos on YouTube platform.
\section{Background}
In this section we describe our problem statement and then discuss data collection and approaches adopted for building the ground truth.

%\subsection{Terminology and Nomenclature}
%A cartoon video $V_c$, in context of its viewing by a child, is said to be \textit{unsafe} if it either contains scenes of graphic nudity or verbal bad language in its dialogues. A comment $C_V$ posted at the cartoon video $V_c$ is said to be \textit{unsafe} if it contains at least one bad term (abusive word or sexually explicit word). Based on the activities with respect to comment posts and video uploads, there can be four types of user behaviours as described in 
%A user is said to be \textit{unsafe content uploader}, in the context of kids, if he or she has uploaded even a \textit{single} indecent video otherwise he or she is \textit{safe uploader}. The \textit{degree of indecency} would vary and would be proportional to the number of indecent videos uploaded out of total videos uploaded. Similarly, a user is said to be \textit{unsafe commenter} if he or she posts even a \textit{single} comment with indecent words in it. The \textit{degree of indecency} would vary and would be proportional to the number of indecent words in comment and number of such comments out of total number of comments.

\subsection{Problem Statement}
More formally, we state the three problems being addressed in this paper as follows.\\
\textbf{Detection Problem}: Given a set of seed cartoon videos $V$, their corresponding set of uploaders $U$ who upload those videos and commenters $C$ who post comments on those videos, the goal is to find (identify) a subset of uploaders $U_{unsafe}$ and commenters $C_{unsafe}$ promoting unsafe content.
\[
f_{identification}(U,C) \rightarrow \{U_{unsafe},C_{unsafe}\}
\]
\textbf{Characterization Problem}: Given a set of unsafe content uploaders $U_{unsafe}$ and safe content uploaders $U_{safe}$, the goal is to find the set of attributes of uploaders that distinguish them from each other to the maximum extent.
\[
f_{characterization}(U_{safe},U_{unsafe}) \rightarrow \{U_{attributes} \dots\}
\]
\textbf{Community Detection Problem}: Given sets of safe and unsafe content promoters, both in terms of video uploads ($U_{safe}$ and $U_{unsafe}$) and commenter posts ($C_{safe}$ and $C_{unsafe}$), the goal is to find (detect) communities within the set of uploaders $U$ (among $U_{safe}$ \& $U_{unsafe}$), within set of commenters $C$ (among $C_{safe}$ \& $C_{unsafe}$) and across the set of uploaders \& commenters. 

\subsection{Data Collection}
YouTube provides a platform for a huge repository of video content and engagements (likes, views , dislikes, comments, etc) around it. In lieu of our focus on content being watched by kids, we restricted ourselves to collection of cartoon videos only. We began searching and collecting videos using top 20 popular cartoon keywords (few indicative ones like mickey mouse, tom and jerry, etc.) programmatically through YouTube Data API.\footnote{https://developers.google.com/youtube/v3/} From the search results, we \textit{randomly} selected 408 videos which formed our seed input. Users who have uploaded these videos, referred as \textit{uploaders}, were found to be 275. For each video, upto 100 latest comments were collected, which commutatively added up to 21,268 comments in total, which were found to be posted by 19,099 unique users, referred as \textit{commenters}. Table \ref{table:data_collected} summarizes the data collected. This seed video dataset was further expanded to 1,178 videos \textit{liked} by these uploaders \& 123,390 videos \textit{liked} by these commenters, 1,293 uploaders \textit{subscribed} by these uploaders \& 131,952 uploaders \textit{subscribed} by these commenters and 1,068 videos \textit{added to playlist} by these uploaders \& 30,148 videos \textit{added to playlist} by these commenters.

\begin{table}[h!]
\centering
\caption{Summary of Data Collected comprising of Seed DataSet collected using Cartoon keywords and an Extended DataSet obtained through engagements of Uploaders and Commenters}
\begin{tabular}{|l|r|}
\hline
\textbf{Description of Data} & \textbf{Count} \\ 
\hline
\multicolumn{2}{|c|}{\emph{Seed Data Set (Cartoon Keywords)}}\\
\hline
Number of Seed Videos & 408 \\
Number of Unique Seed Uploaders & 275 \\
%Number of Decent Videos & 284 \\
%Number of Indecent Videos & 124 \\
Number of Seed Comments & 21,286 \\
Number of Unique Seed Commenters & 19,099 \\
%Number of Decent Comments & 19472 \\
%Number of Indecent Comments &  1814 \\
\hline
\multicolumn{2}{|c|}{\emph{Extended Data Set (Engagements)}}\\
\hline
Number of Extended Videos & 155,784\\
 - Videos \textit{liked by} Uploaders & 1,178\\
 - Videos \textit{liked by} Commenters & 123,390\\
 - Videos \textit{added to playlist} by Uploaders & 1,068\\
 - Videos \textit{added to playlist} by Commenters & 30,148\\
Number of Extended Uploaders & 133,245\\
 - Uploaders \textit{subscribed by} Seed Uploader & 1,293\\
 - Uploaders \textit{subscribed by} Seed Commenters & 131,952\\
\hline  
\end{tabular}
\label{table:data_collected}
\end{table}

More than 80\% of uploaders in the seed dataset contributes only one video in the seed video dataset and similarly around 90\% of commenters in the seed dataset makes only one comment.

\subsection{Building Ground Truth}

All the seed videos that were collected using Cartoon keywords through the YouTube API were manually inspected (annotated) for presence (`1') or absence (`0') of unsafe content in video either in form of graphic nudity or abusive words spoken in dialogues. Out of the total 408 seed videos, 284 were found to be safe and remaining 124 videos were found to be unsafe for viewing by kids. With respect to the comments, we treat each comment as \textit{bag of words} separated by spaces. We compared each word from the comment text with a set of bad words obtained from an online  repository.\footnote{Bad Word List: https://www.cs.cmu.edu/\~biglou/resources/bad-words.txt} Out of 21,268
comments, 1814 comments were found to have atleast one bad word. 

\section{Detection of Unsafe Content Promoters}

In this section we explain our solution to the first problem of detection of unsafe content promoters. Such promoters manifest their malicious activities in two forms, \textit{first}, by uploading indecent videos (unsafe content uploaders) and \textit{second}, by posting indecent comments (unsafe content commenters). 

\subsection{Detecting Unsafe Content Uploaders}
For identifying unsafe content uploaders, we adopt two independent approaches. One approach, referred a \textit{classifier approach} uses features that capture responses of other users in a crowd sourced environment. Features at the level of video, user and comments are used. In the other, referred as \textit{CNN based approach}, features from the actual video are used, details are provided below.

\textbf{Classifier Approach:} In this, we aim to build a classifier model to automatically mark users (uploaders) as unsafe or safe based on three types of features namely video-level, user-level and comment-level which are explained in Table \ref{table:feature_set}, numbers below in brackets refer to number of features used. 

\emph{Video-Level Features (19)}: These capture the cumulative response received from YouTube users in form of view count, comment count, like count and dislike count to the past upto 100 videos uploaded by the uploader. In addition, few other features associated with video included are video duration, length of title, length of description, description to title ratio, jaccard similarity between title and description, bad words found in title and description, number of question marks, hyperlinks and emoticons found in description.

\emph{User-Level Features (9)}: These include the user's profile related features like total number of videos uploaded, total views \& comments received on those videos, number of subscriptions, number of days since joining YouTube and length of title \& description. In addition, there are features associated with YouTube user's Google+ userID namely circledByCount (number of other users in the network of user) and plusOneCount (number of likes received by user) which are also used.\footnote{Google+ API: https://developers.google.com/+/web/api/rest/} 

\emph{Comment-Level Features (6)}: These capture the response of other users through the comments that they have been posting on videos uploaded by the user. Features related to comments used are cumulative likes and replies received by comments, number of positive, neutral and negative sentiment received on comments by the user and number of bad words in comments.

\begin{table}[h!]
\centering
\caption{Exhaustive List of Features (Video-Level, User-Level and Comment-Level) used for building a Classification Model for Detection of Unsafe Content Promoters in YouTube}

\begin{tabular}{|c|}\hline
\textbf{Feature Description} \\ 
\hline
\emph{Video Level Features (computed for upto 100 videos uploaded)}\\
\hline
type of video, number of views received on videos, number of comments \\
posted on videos, number of dislikes obtained on videos, number of likes\\
 received on videos, ratio of number of likes to dislikes, length of title\\
 of videos, length of video description, ratio of description to title of\\
 videos, video duration in seconds, number of days since video was \\
 published, jaccard similarity between words appearing in description \\
 and title, number of bad words in title, number of bad words in \\
 description, number of times `?' appears in description, number of times\\
  hyperlink appears in description, number of emoticons in description, \\
  presence or absence of `18' in video title, common words in title and\\
   description of video. \\ 
\hline
\emph{User Level Features (computed for YouTube Channel \& Google+ userID)}\\
\hline
total number of videos uploaded, total number of views received on all \\
videos uploaded, total number of comments posted on all videos \\
uploaded, number of subscribers of the user, length of user's title,\\
length of user's description, number of days since the user registered\\
  with YouTube, number of Google+ users in the network of the user,\\
  number of likes received from other Google+ users by the user. \\
\hline
\emph{Comment Level Features (computed for upto 50 top comments per video)}\\
\hline
likes received on comments posted on videos uploaded by user, number\\
 of replies made to the comments, number of comments with positive \\
 sentiment, number of comments with negative sentiment, number of \\
 comments with neutral sentiment, number of bad words in comments.\\
\hline
\end{tabular}
\label{table:feature_set}
\end{table}

\begin{figure}[h]
\includegraphics[width=0.5\textwidth]{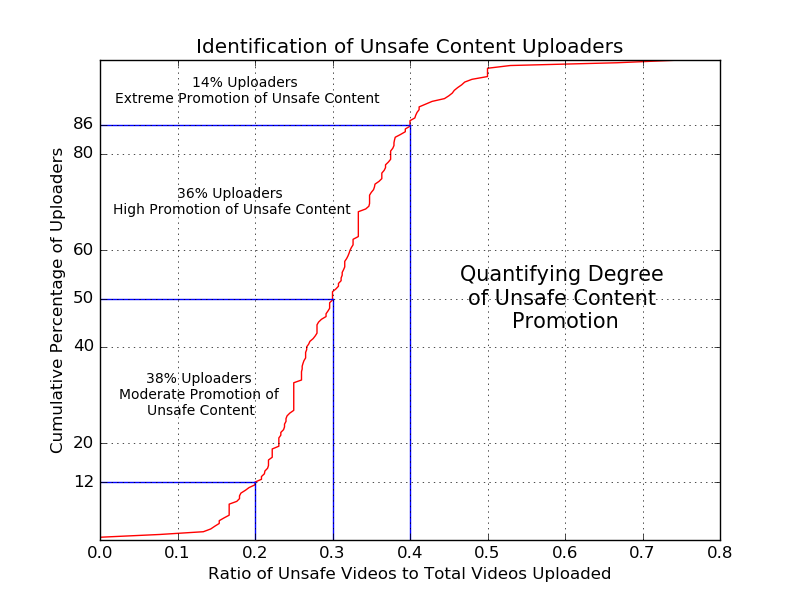}
\caption{Ratio of Indecent Videos to Total Videos by Uploaders obtained after applying Algorithm 1 to upto 50 latest videos uploaded by the seed uploaders}
\label{fig:ratio}
\end{figure}

We apply a number of conventional classifiers (\ref{table:classifier_results}) on these features which are computed for the seed videos dataset as input with 80:20 training to test split and found that \textit{random forest} classifier performs the best with respect to accuracy in predicting the class (safe or unsafe) of an input video. The classifier, thus developed, was used to predict the safety (classs of video whether decent or indecent) of upto 50 latest videos uploaded for each of the unique uploader in the seed dataset. Algorithm 1 describes the entire process of identifying unsafe content uploaders using video classifier. After applying Algorithm 1, it was observed that there are some users (uploaders) who have uploaded both decent and indecent videos, a cumulative distribution frequency (CDF) plot of proportion of  indecent (unsafe) videos to total videos uploaded by users is depicted in Figure \ref{fig:ratio}. This plot helps us in choosing a thresholds to mark an uploader as unsafe or safe based on the amount of unsafe (indecent) videos uploaded by them. \nolinebreak

\textbf{Algorithm 1:} Detect-Unsafe-Promoter-By-Classifier ($V$)
\hrule
\begin{algorithmic}[1]
\renewcommand{\algorithmiccomment}[1]{#1}
\STATE $V \leftarrow {set\ of\ seed\ videos}$
\FORALL{$v \in V_u$}
\STATE $V_f \leftarrow {find\_features\_at\_level (video,user,comment)}$
\ENDFOR
\STATE $C_v \leftarrow video\_classifier (V_f)$
\STATE $U_V \leftarrow {set\ of\ unique\ uploaders\ of\ V}$
\FORALL{$u \in U_V$}
\STATE $V_u \leftarrow find\_past\_uploaded\_videos(u)$
\FORALL{$v \in V_u$}
\STATE $v_{label} \leftarrow C_v.predict(v)$
\ENDFOR
\ENDFOR
\end{algorithmic}
\hrule 

\begin{table}[h!]
\centering
\caption{Results of Classifiers for Predicting Video Safety}
\begin{tabular}{|l|l|c|c|c|}\hline
\textbf{Classifier Name} & \textbf{Feature List} & \textbf{Precision} & \textbf{Recall} &  \textbf{Accuracy}\\ 
\hline
\multirow{4}{*}{Random Forest} 
  & Video-Level & 83.3 & 66.6 & 82.9 \\
  & User-Level & 62.5 & 47.6 & 65.3 \\
  & Comment-Level & 60.0 & 60.0 & 70.7 \\
  & All Features & \textbf{88.8} & \textbf{76.1} & \textbf{85.7} \\
  \hline
  \multirow{4}{*}{K-Nearest Neighbor} 
  & Video-Level & 57.1 & 13.3 & 64.6 \\
  & User-Level & 75.0 & 28.5 & 65.3\\
  & Comment-Level & 80.0 & 26.6 & 70.7\\
  & All Features & 66.6 & 47.6 & 67.3 \\
  \hline
  \multirow{4}{*}{Decision Tree} 
  & Video-Level & 77.2 & 56.6 & 78.0 \\
  & User-Level & 46.4 & 43.3 & 60.9 \\
  & Comment-Level & 44.8 & 43.3  & 59.7 \\
  & All Features & 68.1 & 71.4 & 73.4 \\
  \hline
\end{tabular}
\label{table:classifier_results}
\end{table}

In reference to the Figure \ref{fig:ratio}, we detect unsafe content uploaders and categorize them into extreme, high and moderate content promoters depending upon the degree of unsafe content in form of child unsafe videos uploaded by them in the past. 

\textbf{CNN based approach}: In this approach we use the video frames and apply the concept of deep learning through Convolutional Neural Networks (CNN). The approach is based on the hypothesis that unsafe video would have transitions (either abrupt or gradual) arising due to the scenes which are depicting nudity in the video. Process of finding these transitions in a video is explained next. Key frames involving abrupt transitions (based on the visual similarity of neighbouring frames of the video) from each video are first extracted. The descriptive efficiency of both SURF (Speeded Up Robust Features) and HSV (Hue, Saturation, Value) histograms descriptors are exploited for assessing frame similarity. More specifically, these abrupt transitions are initially detected between successive video frames where there is a sharp change and thereafter the calculated scores are further analysed for the identification of frame sequences where a progressive change of the visual content takes place and in this way gradual transitions are detected as well. Finally, a post-processing step is performed to identify outliers due to object/camera movement and flash lights. We implement a CNN to detect the dissimilar image in a video frame. CNN comprises of an input layer, two convolutional layers, two sampling layers, five kernels each of size 3x3 in convolutional layer, mean filter of 2x2 in sampling layer, one fully connected layer and target output layer. The five kernels use five different feature maps namely gray, gradient-X, gradient-Y and last two kernels have randomly initialized values.
%\emph{Back Propagation}: Following steps are performed. (1) An Error Matrix $E$ is  obtained  by finding difference between values of neurons in output layer and fully connected layer. (2) Since there are 5 kernels in convolutional layers, each will have 5 different feature maps. Therefore, in a fully connected layer of Object Recognition CNN, we a total of 18x18x5 neurons (feature maps). (3) Mean error $E_{mean}$ for each feature map is calculated and is used as a threshold for success or failure. (4) Error of each neuron is propagated backwards and thus weights are updated accordingly. Back-propagation comes to a halt when error is less than 0.0003 or number of epochs is 64 for training.
This approach was found to detect unsafe video with 74\% accuracy which is less than the accuracy obtained using the earlier approach. It was further observed that this approach is computationally expensive since it involves video processing on frame-by-frame basis. Due to low accuracy and high computational cost, this approach was not pursued further.
 
\subsection{Detecting Unsafe Content Commenters}
Since there is no way in which a user's past comments could be retrieved through the YouTube API, so a user is marked as unsafe (indecent) commenter if he or she has posted at least one comment which carries indecent words. Out of 21268 comments, 1814 comments were found to have atleast one bad word which were posted by 1755 unique users, all such users were marked as unsafe content commenters for the purpose of our work. 

\section{Uploader Characterization Study}
After having identified unsafe content uploaders, in this section, we characterize these uploaders with respect to the popularity and engagements received on the videos uploaded by them. The goal is find distinguishing features which separates unsafe content promoters from those spreading safe content for kids. 
%In the context of YouTube, uploaders are referred as \textit{channels} each of which has a unique \textit{channel ID}. It can either be managed by a google account or through a google+ page, details can be referred here.\footnote{https://support.google.com/youtube/answer/4642409?hl=en}

\subsection{Popularity Factors}
\begin{figure*}[!t]
\centering
\subfloat[Subscription Count]{\includegraphics[width=1.7in]{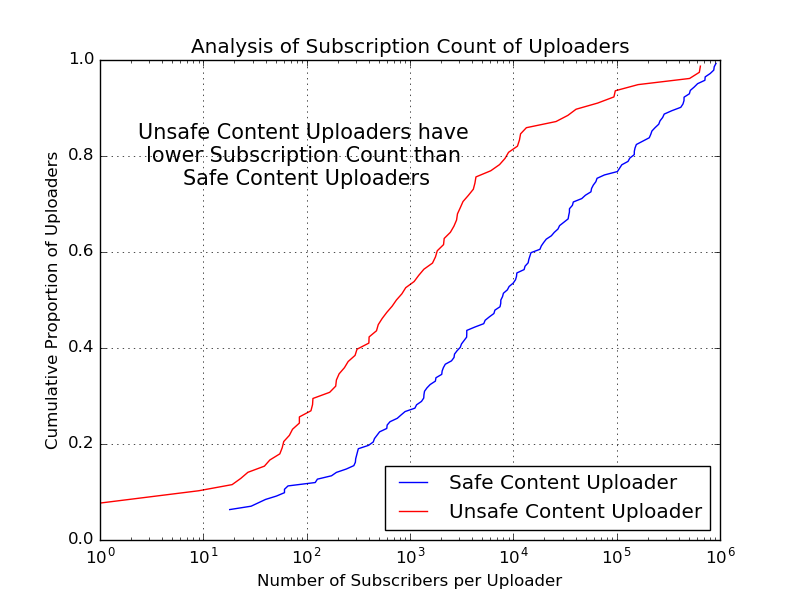}%
\label{fig_subscribe_count}}
\hfil
\subfloat[Viewership]{\includegraphics[width=1.7in]{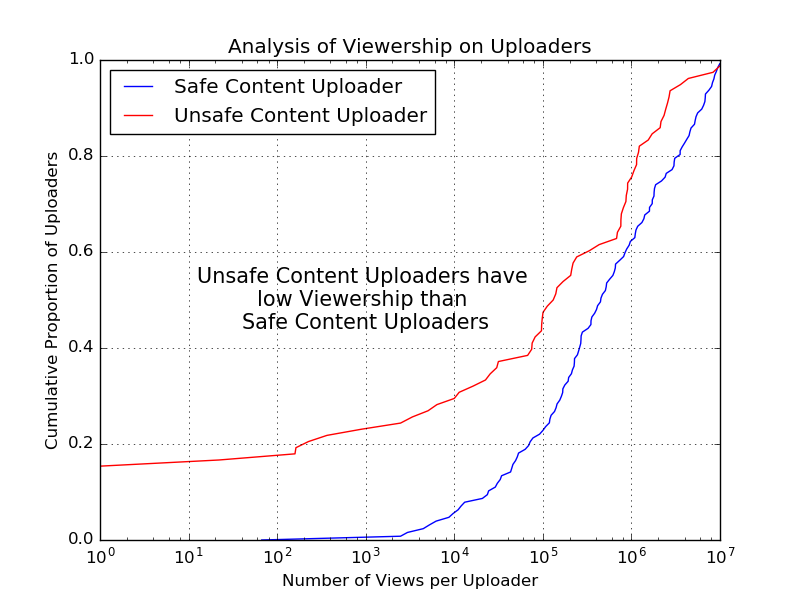}%
\label{fig_viewership}}
\hfil
\subfloat[Google+ Network]{\includegraphics[width=1.7in]{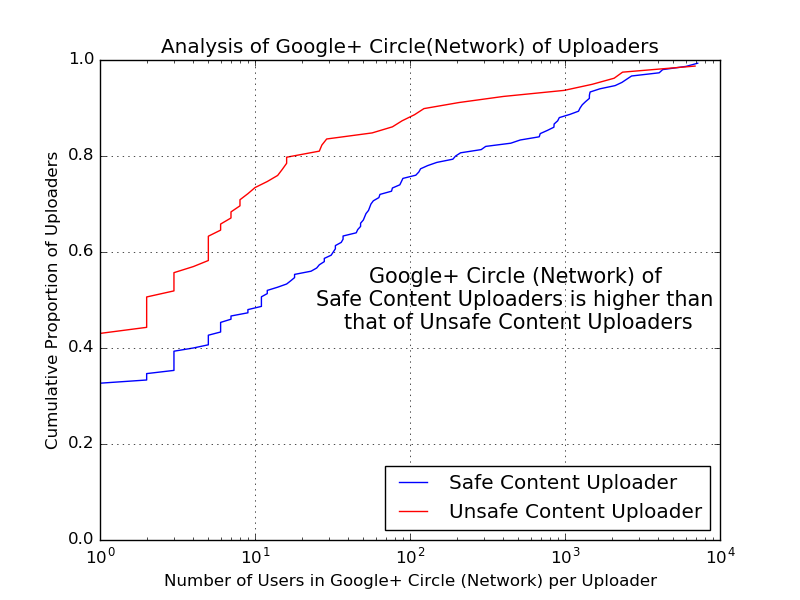}%
\label{fig_gplus_network}}
\hfil
\subfloat[Uploaded Video Count]{\includegraphics[width=1.7in]{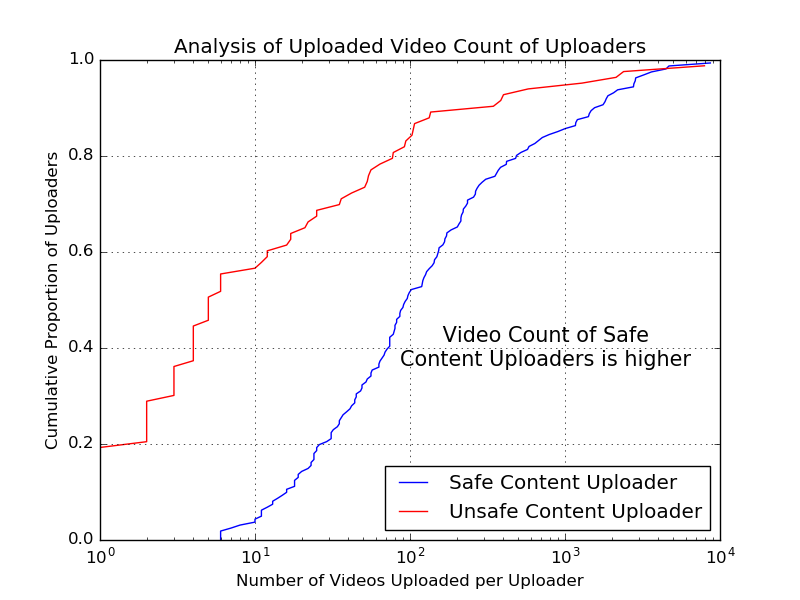}%
\label{fig_video_count}}
\caption{Characterization of Safe and Unsafe Content Uploaders based on their Popularity Metrics}
\label{fig:popularity}
\end{figure*}

Popularity is an important factor to characterize and distinguish unsafe content uploaders from safe content uploaders. In context of YouTube, popularity can be gauged by various factors whose characterization study is depicted in Figure \ref{fig:popularity} comprising subscriber count, circledByCount, total videos count and total view count received on all the videos uploaded by the user. Number of subscribers to a YouTube channel is one of the key indicators of popularity of the uploader. Figure \ref{fig_subscribe_count} depicts that unsafe content uploaders have a comparatively less number of subscribers than safe content uploaders which is suggestive of the fact that users subscribe more to those spreading safe content.

\begin{figure*}[!t]
\centering
\subfloat[Comment Count]{\includegraphics[width=2.0in]{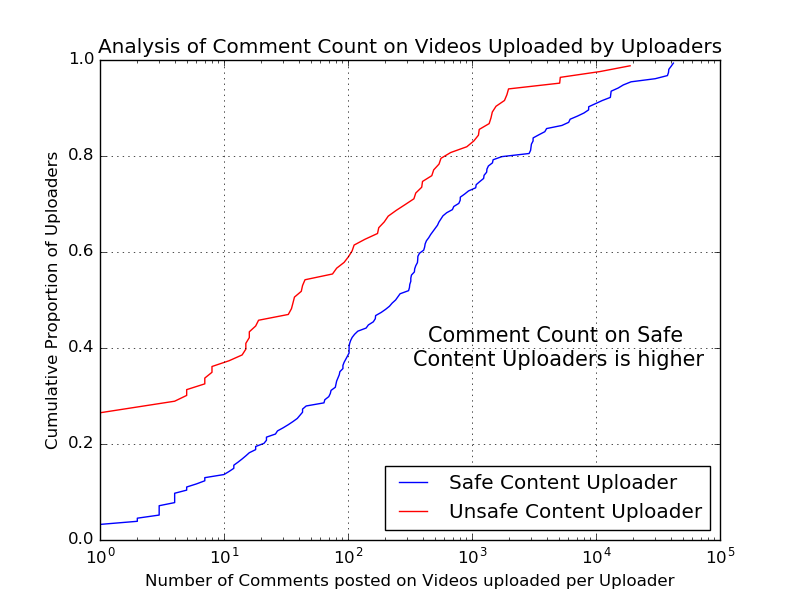}%
\label{fig_cmt_count}}
\hfil
\subfloat[Like Count]{\includegraphics[width=2.0in]{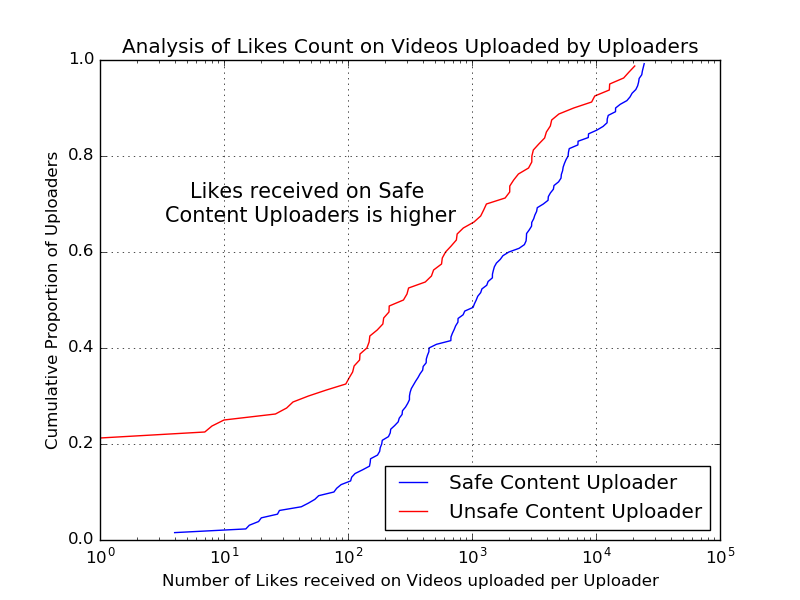}%
\label{fig_like_count}}
\hfil
\subfloat[Dislike Count]{\includegraphics[width=2.0in]{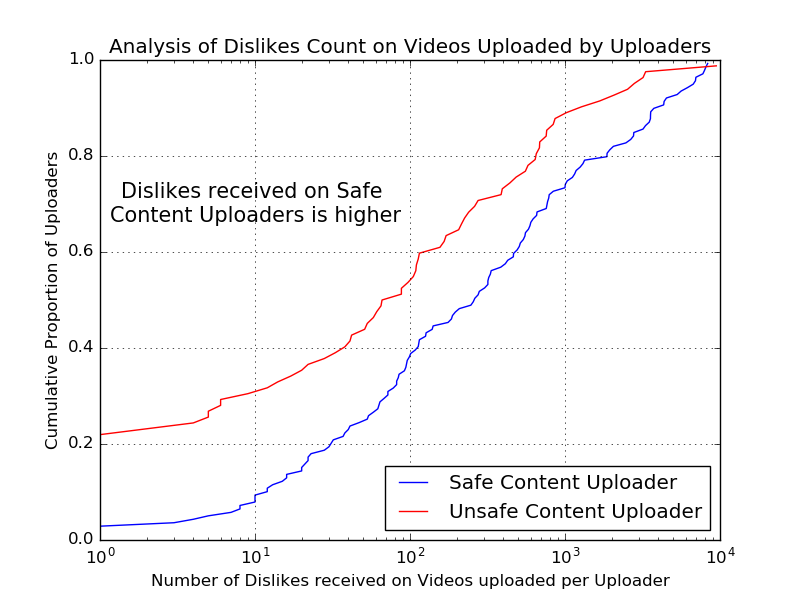}%
\label{fig_dislike_count}}
\caption{Characterization of Safe and Unsafe Content Uploaders based on their Engagement Metrics}
\label{fig:engagement}
\end{figure*}

Number of videos uploaded by safe content uploaders are more than those uploaded by unsafe content uploaders as depicted in Figure \ref{fig_video_count} which indicates that users spreading unsafe content are not inclined towards maintaining a long standing profile through their video content on YouTube. Viewership trends for safe content uploaders is higher than that of unsafe content uploaders as per Figure \ref{fig_viewership}, which could perhaps be due to the fact the unsafe content uploaders upload less videos as well. Finally, each uploader represented as a YouTube Channel is typically managed through a Google+ account whose network is studied in Figure \ref{fig_gplus_network} which shows that unsafe content uploaders have a lessor users in their network (circle) as compared to those spreading safe content which is in coherence with other popularity metrics studied.

\subsection{Engagement Factors}
To further understand the key distinguishing features that separates safe content uploaders from unsafe content uploaders, we study engagements. For capturing engagements three factors namely comments, likes and dislikes received on past videos posted by uploaders are compared in Figure \ref{fig:engagement}. 

Comments, likes and dislikes received on videos uploaded by the safe and unsafe content uploader are depicted in Figure \ref{fig_cmt_count}, Figure \ref{fig_like_count} and Figure \ref{fig_dislike_count}, respectively.  In all of them the engagements received by safe content uploaders is higher than unsafe content promoters suggesting that users (who are typically kids) engage more with those spreading safe content.

\section{Community Detection}
After having characterized the popularity and engagement of safe and unsafe content uploaders, in this section, we perform network based analysis to understand the underlying interconnected structures among the uploaders and commenters on YouTube spreading safe and unsafe content. The purpose is to detect communities among the safe and unsafe content promoters (both in respect to uploaders and commenters). These communities would help us in understanding whether safe content promoters are closely knit with the unsafe content promoters. If it is so then such community structures could end up driving kids moving (or viewing) from safe content promotion space to a space comprising of unsafe content promoters, as would be argued in this section later. Three ways of detecting such communities are explained next. 

\subsection{Video Network Analysis}
In our \textit{first approach} for detecting communities, \textit{related videos} feature of YouTube is leveraged to find a list of videos that are suggested for an given input video. We use this feature to find video-video communities based on these suggested (related) videos by using YouTube API.\footnote{https://developers.google.com/youtube/v3/docs/search/list} We construct a directed graph comprising of videos as nodes, in which a directed edge between two videos $v_i$ and $v_j$ is formed if $v_j$ is found to be \textit{related to} $v_i$. Algorithm 2 constructs the graph by taking the seed videos dataset as input. For each video, algorithm finds top 10 related videos. If the related video is from among the set of seed videos dataset, then an edge is added in the graph. Procedure $update\_transitions$ counts all possible transitions (edges) between safe and unsafe videos in the graph. From the set of 408 seed videos given as input, it was found that there are 262 nodes and 630 edges formed between them.

\textbf{Algorithm 2} Video-Video-Community-Detect ($V$)
\hrule
\begin{algorithmic}[1]
\renewcommand{\algorithmiccomment}[1]{#1}
\STATE $V \leftarrow {set\ of\ seed\ video\ dataset}$
\FORALL{$v \in V$}
\STATE $G \leftarrow add\_node(v) $
\ENDFOR
\STATE $th \leftarrow 10$
\FORALL{$v \in V$}
\STATE $v_{related} \leftarrow find\_related\_videos(v,th)$
\FORALL{$v_r \in v_{related}$}
\IF{$v_r$ is a video in $V$}
\STATE $G \leftarrow add\_edge(v,v_r) $
\STATE $update\_transitions(v,v_r)$
\ENDIF
\ENDFOR
\ENDFOR
\end{algorithmic}

\begin{figure}[h]
\centering
\includegraphics[width=0.4\textwidth]{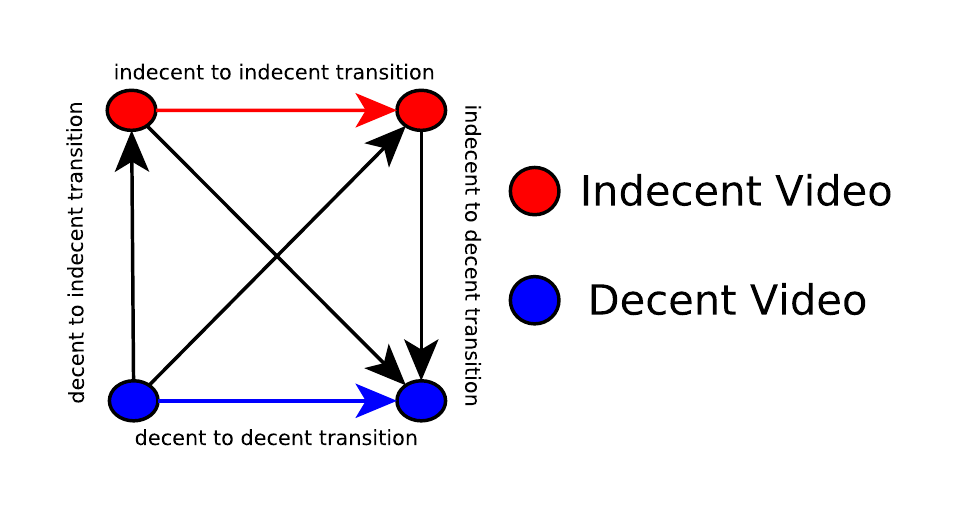}
\caption{Possible Video-Video Transitions based on Video Suggestion feature of YouTube}
\label{fig:relatedVideoDiagram}
\end{figure}

Community detection algorithm was run based on the work of Vincent et. al. \cite{Vincent} provided by Gephi.\footnote{https://gephi.org/} A total of 31 communities with Louvain Modularity value of 0.807 were found indicating that communities are fairly well connected among themselves. Figure \ref{fig:relatedVideoDiagram} depicts the four set of transitions among decent and indecent videos that we intend to capture through this video-video graph. Table \ref{table:5} depicts the distribution of 630 edges (transitions) among decent and indecent videos, thereby summarizing the results for all possible transitions. 

The implication of this analysis is to suggest that while safe to unsafe transition is quite a rare (9 and 14 in video-video and uploader-uploader, respectively) phenomena, the unsafe to unsafe transition is quite prevalent in the case of video-video transition where it happens 117 times. This indicates that once a user (typically a child watching cartoons) moves from decent (safe) video space to indecent (unsafe) video space, he or she would remain in it due to video suggestions. The idea of video-video transitions is extended to the uploaders as well and an uploader-uploader graph is constructed on similar lines and uploader-uploader transitions are also mentioned in Table  \ref{table:5}.

\begin{table}[h!]
\centering
\caption{Video-Video and Uploader-Uploader Transitions among the Communities formed based on Video Suggestions Feature}
\begin{tabular}{|l|c|c|}\hline
\textbf{Description} & \textbf{Video-Video} & \textbf{Uploader-Uploader} \\ \hline
Number of Nodes & 262 & 92\\
\hline
Number of Edges & 630 & 114\\
\hline
Number of Communities & 31 & 21\\
\hline
Modularity & 0.807 & 0.816\\
\hline
Safe to Safe Transition & 498 & 94\\
Safe to Unsafe Transition & 9 & 14\\
Unsafe to Safe Transition & 6 & 5\\
Unsafe to Unsafe Transition & 117 & 1\\
\hline
\end{tabular}
\label{table:5}
\end{table}

\subsection{Uploader-Commenter Network Analysis}

Besides being a platform for video sharing, users on YouTube post comments on videos. We leverage this engagement on YouTube to construct uploader-commenter undirected graph by considering seed uploaders and seed commenters as nodes (users). An edge between two users is formed if one user comments on video uploaded by another user. Due to lack of space, the algorithm for graph construction is not mentioned. In order to capture only the significant relationships among commenters and uploaders, we considered only those commenters who have posted at least 4 comments. With this change, algorithm resulted in a undirected graph comprising of 1942 nodes (of commenters and uploaders) and 2373 edges among them. Running community detection algorithm on Gephi resulted in 147 communities with Louvain Modularity value of 0.973 again indicating well connected communities in the graph. Such a graph between uploaders (safe and unsafe) and commenters (safe and unsafe) brings forth many interesting observations. One of them being that even though an uploader may be quite decent (safe) and has posted a decent (safe) video, the fact that a commenter posts an indecent (unsafe) comment, could effectively undermine the uploader's safe behavior. Table \ref{table:7} highlights this observation in which 733 transitions (edges) are found from indecent (unsafe) commenters on decent (safe) uploaders. 

\begin{table}[h!]
\centering
\caption{Engagements between Commenter Uploader obtained through Comments made on Videos uploaded by Uploaders}
\begin{tabular}{|l|c|}\hline
\textbf{Transition Type} & \textbf{•}{Count} \\ \hline
Safe Commenter to Safe Uploader & 396 \\
Unsafe Commenter to Safe Uploader & 733 \\
Safe Commenter to Unsafe Uploader & 156 \\
Unsafe Commenter to Unsafe Uploader & 1088 \\
\hline
\end{tabular}
\label{table:7}
\end{table}

\subsection{Uploader-Commenter Behavioural Analysis}
Commenters and uploaders are also connected with each other through various other forms of behaviour engagements. Three types behavioural engagements studied are `who \textit{likes} whom', `who \textit{subscribes} whom' and `who adds whom to \textit{playlist}'. Communities among safe and unsafe content promoters (uploaders and commenters) formed through the manifestation of these behaviours are investigated. It may be noted here that YouTube API doesn't provide a direct function to find list of users who have liked a particular video nor can we find a list of users who have subscribed a particular user. In order to find this information, we take a user as input and first find all the videos liked, all the other users subscribed and all the videos added in the playlist by this user. We then take these results and do a \textit{reverse lookup} to find if any of these users and videos exist in our seed dataset in order to establish a behavioural relationship. The detailed procedure for this approach is not mentioned due to space considerations. With 275 seed uploaders and 19099 seed commenters, Tables \ref{table:likes_stats}, \ref{table:subscriber_stats} and \ref{table:playlist_stats} summarize the results of behavioural engagements among them based on likes, subscription and playlist. 

\begin{table*}
\centering
\setlength\tabcolsep{2pt}

\begin{minipage}{0.3\textwidth}
\centering
%\tablewidth=\textwidth

%\begin{minipage}{.2\textwidth}
%\centering
\caption{Who `likes' Whom Statistics}
\begin{tabular}{|l|r|}\hline
\textbf{Description of Behaviour} & Count \\ \hline
Number of Seed Uploaders & 275 \\
Number of Seed Commenters & 19,098 \\
\hline
Total Videos liked by Uploaders & 1,178 \\
 - Likes on Videos by Self & 186 \\
 - Likes on Videos by Other Uploaders & 24 \\
 - Likes on Videos by Commenters & 4 \\
 - Likes on Videos by Other Users & 964 \\
\hline
Total Videos liked by Commenters & 123,390 \\
 - Likes on Videos by Self & 1,111 \\
 - Likes on Videos by Other Commenters & 604 \\
 - Likes on Videos by Uploaders & 2,057 \\
 - Likes on Videos by Other Users & 119,618 \\
\hline
\end{tabular}
\label{table:likes_stats}
\end{minipage}%
\hspace{8mm}
\begin{minipage}{0.3\textwidth}
\centering
%\tablewidth=\textwidth

\caption{Who `subscribes' Whom Statistics}
\begin{tabular}{|l|r|}\hline
\textbf{Description of Behaviour} & Count \\ \hline
Number of Seed Uploaders & 275 \\
Number of Seed Commenters & 19,098 \\
\hline
Total Subscription by Uploaders & 1,293 \\
 - Subscription to Other Uploaders & 27 \\
 - Subscription to Commenters & 3 \\
 - Subscription to Other Users & 1,263 \\
\hline
Total Subscription by Commenters & 131,952 \\
 - Subscription to Other Commenters & 617 \\
 - Subscription to Uploaders & 2,086 \\
 - Subscription to Other Users & 129,249 \\
\hline
\end{tabular}
\label{table:subscriber_stats}
\end{minipage}
\hspace{2mm}
\begin{minipage}{0.3\textwidth}
\centering
%\tablewidth=\textwidth

\caption{Who adds Whom to `playlist' Statistics}
\begin{tabular}{|l|r|}\hline
\textbf{Description of Behaviour} & Count \\ \hline
Number of Seed Uploaders & 275 \\
Number of Seed Commenters & 19,098 \\
\hline
Total Videos in playlist of Uploaders & 1,068 \\
 - Self Videos & 757 \\
 - Videos of Other Uploaders & 1 \\
 - Videos of Commenters & 2 \\
 - Videos of Others Users & 308 \\
\hline
Total Videos in playlist of Commenters & 30,148 \\
 - Self Videos & 2,148 \\
 - Videos of Other Commenters & 117 \\
 - Videos of Uploaders & 440 \\
 - Videos of Other Users & 27,407 \\
\hline
\end{tabular}
\label{table:playlist_stats}
\end{minipage}

\end{table*}

\subsubsection{Likes Network Community}
Self likes of users (uploaders and commenters) refer to the likes made to videos uploaded by themselves. As evident from Table \ref{table:likes_stats}, 24 out of 275 uploaders (nearly 8\%) like videos uploaded by other uploaders whereas only 604 commenters are connected out of 19098 commenters through likes. Among the others users (referred as 'Other Users' because these YouTube channel IDs neither belong to uploaders nor commenters in our dataset) whose videos are liked by uploaders and commenters, there are 262 common other users which are liked by both uploaders and commenters suggesting similar interest between them.

\subsubsection{Subscriber Network Community}
Subscribers are other YouTube channels that a user typically subscribes to on the basis on his or her interest. In this section, we collect and analyze all the subscribers (YouTube channels) subscribed by uploaders and commenters. Table \ref{table:subscriber_stats} refers to the related data and statistics related to the `who subscribes whom' graph and analysis around it follows next. There are no self-subscribers since self subscription would not make any sense. Around 10\% of uploaders are connected with each other through subscriptions while only 617 out of 19098 of commenters are connected with each other. Commenters subscribe to more number of uploaders while uploaders subscribing to commenters is very rare. 

\subsubsection{Playlist Network Community}
YouTube provides a feature of adding videos that a user would typically want to view occasionally and combine such videos into a playlist. Give a user (YouTube channel ID which is either uploader or commenter), we can obtain details of all the playlist of that user. Using this approach, playlists created by all the uploaders and commenters are collected and analyzed, Table \ref{table:playlist_stats} presents a brief summary of the same. Among uploaders, most (757 out of 1068) of the connections are due to uploaders adding their own uploaded videos in their playlist. Adding videos of other uploaders and commenters in their playlist is rare. In contrast among commenters, very few (2148 out of 30148) videos uploaded by commenters are in their playlist, most of the videos from other users are in their playlist. Commenters and uploaders adds videos from 67 common users belonging to category of the other users. 

\subsubsection{Combined Network Community}
\begin{figure}[h]
\includegraphics[width=0.5\textwidth]{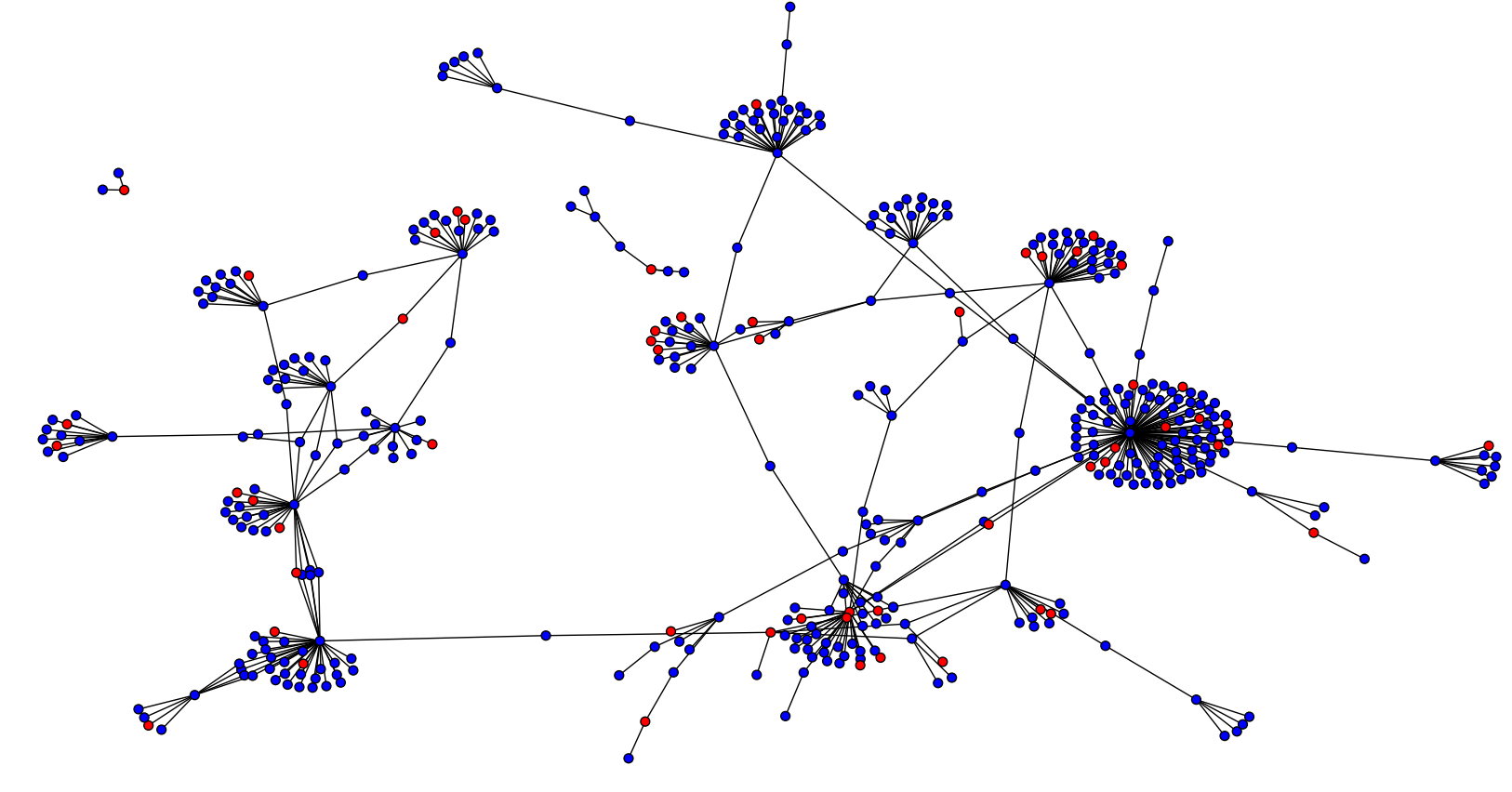}
\caption{Snapshot of Graph comprising of Safe ('blue') and Unsafe ('red') Content Promoters depicting the large closely knit community between them.}
\label{fig:combined_network}
\end{figure}

To understand a combined behavioural analysis contributed through likes, subscription and playlist, we constructed a graph comprising of all uploaders and commenters as nodes. An edge is established between the nodes if they are related either through likes, subscription or playlist. Figure \ref{fig:combined_network} depicts one of the largest communities formed in this graph. Nodes in `red' are the unsafe content promoters (either uploaders or commenters) and those in `blue' are safe content promoters. One can clearly visualize closely knit communities among safe and unsafe content promoters. Since these communities are built on behaviours, it further reinforces the concern expressed earlier that a naive user (kid) watching content from a safe content promoter could easily move into the space of unsafe content promoters.  

\section{CONCLUSION}
In this work, we were able to successfully detect child unsafe content and its promoters on YouTube. Further, we characterized such promoters with other users in terms of popularity and engagement. Finally, we found close communities of child unsafe promoters with other users.

% sigproc.bib is the name of the Bibliography in this case
% ACM needs 'a single self-contained file'!
%\balancecolumns

%Appendix A
%\balancecolumns % GM June 2007
% That's all folks!
\end{document}